# A Latent Feature Analysis based Approach for Spatio-Temporal Traffic Data Recovery

Yuting Ding

*School of Computer Science and Technology,*

*Chongqing University of Posts and Telecommunications*

*Chongqing Institute of Green and Intelligent Technology, Chinese Academy of Sciences*

*Chongqing School, University of Chinese Academy of Sciences*

Chongqing 400714, China

s200231024@stu.cqupt.edu.cn

Di Wu

*College of Computer and Information Science,*

*Southwest University,*

Chongqing 400715, China,

wudi.cigit@gmail.com

*Abstract*—Missing data is an inevitable and common problem in data-driven intelligent transportation systems (ITS). In the past decade, scholars have done many research on the recovery of missing traffic data, however how to make full use of spatio-temporal traffic patterns to improve the recovery performance is still an open problem. Aiming at the spatio-temporal characteristics of traffic speed data, this paper regards the recovery of missing data as a matrix completion problem, and proposes a spatio-temporal traffic data completion method based on hidden feature analysis, which discovers spatio-temporal patterns and underlying structures from incomplete data to complete the recovery task. Therefore, we introduce spatial and temporal correlation to capture the main underlying features of each dimension. Finally, these latent features are applied to recovery traffic data through latent feature analysis. The experimental and evaluation results show that the evaluation criterion value of the model is small, which indicates that the model has better performance. The results show that the model can accurately estimate the continuous missing data.

*Keywords—intelligent transportation systems, latent feature analysis, spatio-temporal correlation, data recovery.*

## I. Introduction

With the steady growth of travel demand, roads in cities all over the world are increasingly congested, but for economic and environmental reasons, this problem can no longer be solved by building new roads [1]. Therefore, optimizing the existing traffic network has increasingly become a more ideal alternative to managing traffic congestion [2]. Intelligent Transportation Systems (ITS) play an important role in optimizing existing transportation systems. As a key input to ITS, traffic data collected in real time can optimize traffic networks, such as route planning and driver assistance systems. With the development of data acquisition technology, traffic data collected from multiple sources, such as ring detectors, GPS and video sensors have become increasingly important [3, 4].

Although traffic speed information is important, unfortunately, large amounts of data are often lost due to various failures in data collection and recording systems, such as failed cycle detectors, failed cycle amplifiers, and failed signal communication and processing equipment [5]. Fortunately, ITS usually exhibit intrinsic spatial-temporal patterns. It is still an open issue of making full use of spatial-temporal traffic patterns to improve recovery performance [6, 7]. For example, two GPS data with close locations tend to have similar traffic features in a small time range(e.g., one second). On this basis, many approaches about the low-rank matrix approximation(LRMA) [8-11] are proposed to address the problem of discovering traffic patterns from partially

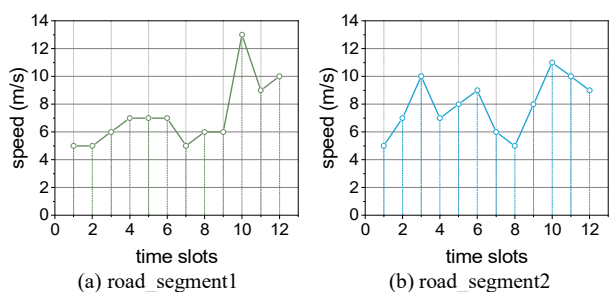

Fig.1. The data distribution of real Shenzhen traffic data.

observed data and use them to recovery the miss data accurately.

The principle of LRMA is to use the incomplete traffic data into a matrix, where the vehicle is constructed to the rows and the columns represent time slots. Then, the LRMA model learns its observed ones only to estimate the missing data of the incomplete matrix. In road networks, the accuracy of LRMA estimation of missing values is higher than other estimation methods. However, classical LRMA models [12]–[18] may overlook the specificity of recorded traffic, as this approach simply treats traffic data as a whole and computes missing values by imposing low-rank constraints globally. For example, [19] studied several temporal imputation such as Historical Average and Surrounding Time Periods. They concluded that missing traffic data of a single location as time series and then impute missing data based on the relationship identified from historical past-to-future data pairs. In fact, trends in traffic flow are not entirely consistent across the sample, depending on different road segments and different working days. This indicates that the degree of linear dependence between traffic flows varies greatly. However, the current LRMA-based imputation ignores these subtle but crucial distinctions and will inevitably suffer.

**Example 1.** We choose the actual vehicle speed dataset from the city of Shenzhen. Fig. 1 depicts the data distribution of the average vehicle speed over time in two nearby Shenzhen locations at 17:00 on October 24, 2014. From Fig. 1, it is clear that while the distribution of vehicle speeds on individual road segments varies over time, that of vehicle speeds on adjacent road segments does not alter significantly. At the same time, it is also discovered that the speed of vehicles on a certain stretch of road is nearly same. Besides, it is also discovered that the speed of vehicles on a certain stretch of road is nearly the same. Using Fig. 1(a) as an illustration, the average speed of the first and second time slices is 5, and the average speed of Fig. 1(a) and Fig. 1(b) in the first time slice is similarly 5, as shown in Fig. 1.

Example 1 shows that there is substantial spatio-temporal correlation in the ITS data, therefore we can create an LRMA model that takes this into account and enhance the model's ability to recovery traffic data. we propose in this paper a latent feature analysis(LFA) method combined with spatial-temporal to recovery traffic data(RTD), named LFA-RTD. Specifically, the proposed methods are significant from in the following aspects:

(1) An LFA-RTD model is proposed. It can accurately recovery missing data based on partially observed data in ITS.

(2) Theoretical analyses are provided for the proposed LFA-RTD model.

(3) Algorithm design for the proposed LFA-RTD model.

## II. RELATED WORK

Numerous sophisticated LRMA models have so far been put out, such as ones that are bias-based [20], non-negativity-constrained [21], dual-regularization-based [22], and neighborhood-and-location integrated [23]. They all employ an L2 norm-oriented Loss that is very sensitive to outliers, albeit they differ from one another in terms of objective functions or learning methods [24-27].

Early traffic data recovery models largely relied on temporal patterns and hardly ever made use of the spatial structure of the data. The most basic approach is Historical Average, which replaces missing numbers with average values from similar time periods in the past [28]. A Bayesian network was used by [29] to learn the probability distribution from observed data and to impute missing data using the best fit. Using the daily periodicity and interval fluctuation of traffic data, [30] established a method known as probabilistic principle component analysis (PPCA). Recent studies use geographical information to recreate lost data. Support vector regression (SVR) and a genetic algorithm were used by [31] to capture the temporal and spatial correlations in the traffic network. Using the geometry of road sections, [32] suggested a modified k-nearest neighbor (KNN) approach that imputes missing data. [33] proposed a spatial context sensing model that uses data from nearby sensors to recreate traffic data. These models show that spatial information is useful for the imputation of traffic data. But instead of fully exploiting global spatio-temporal information, they have concentrated on using local spatial information from nearby areas.

Numerous deep neural network models have been created to handle the traffic data imputation challenge as a result of recent developments in deep learning. [34] used generative models made of bidirectional RNNs to fill in the gaps in text data. To address the issue of missing data, [35] proposed a neural network model known as denoising stacked auto-encoder. Although these methods show that deep learning is excellent for data imputation, they hardly take geographical information into account. In order to take advantage of spatial correlations, [36] proposed a multi-view learning strategy that uses SVR to capture spatial dependencies and LSTM to capture temporal dependencies. [37] suggested a convolution recurrent auto-encoder for missing data imputation, modeling spatial correlations using multi-range CNNs. The non-Euclidean linkages on irregular road networks are not considered, even if CNNs perform well for Euclidean correlations in grid-structured data (such as photographs).

## III. PRILIMINARIES

### A. Symbols and Notations

TABLE I. SYMBOL ANNOTATIONS

| Symbol | Explanation |
|---|---|
| $H$ | Original incomplete data matrix, where each row corresponds to a vehicle node and each column corresponds to a time slot. |
| $M, N$ | The number of rows and columns of $H$ respectively. |
| $L$ | The Laplacian matrix |
| $W$ | Weight matrix |
| $M_{i,j}, N_{i,j}$ | $M$ and $N$'s element at the $i$-th row and $j$-th column. |
| $D$ | Temporal differential matrix |
| $d$ | Latent factor dimension |
| $X, Y$ | Latent factor matrices. |
| $x_i, y_j$ | $i$-th row-vector and $j$-th row-vector of $X$ and $Y$ |
| $T$ | The maximum iteration number. |
| $t$ | The $t$-h iteration, $t \in \{1, 2, …, T\}$. |
| $\lambda$ | The regularization parameter. |
| $\eta$ | Learning rate. |
| $\beta_1, \beta_2$ | Hyper-parameters of controlling the effects of spatial and temporal smoothness, respectively. |

### B. LFA-Based Data Recovery in ITS

**Definition 1.** Given a ITS data with $M$ rows and $N$ columns, $H^{M \times N}$. which each element $h_{i,j}$ describes $H$'s element at the $i$-th row and the $j$-th column. Where each row corresponds to a vehicle and each column represents a time slot. Let $H_K$ and $H_U$ denote their known and unknown entry sets, respectively.

In the real situation in most ITS data, the position of the vehicle is dynamic, and is affected by external factors.

**Definition 2.** Given $H$ and $d$, an LF model manages to Matrix complement $X^{M \times d}$ and $Y^{N \times d}$ for $H = XY^T$ [38] based on $H_K$, only with $d \ll \min\{M, N\}$.

The original data matrix can be represented as $H$ where $M$ is the amount of vehicles and $N$ is the number of periods, the objective function can be modeled as:

$$\varepsilon(M,N) = \frac{1}{2} \| J \circ (H - XY^T) \|_F^2 \qquad (1)$$

Where ∘ denotes the Hadamard product performing the elementwise multiplication between two matrices, $J$ is the sampling matrices. The objective function can be modeled as follows:

$$\varepsilon(X,Y) = \frac{1}{2} \| J \circ (H - XY^T) \|_F^2 + \frac{\lambda}{2} (\|X\|_F^2 + \|Y\|_F^2) \qquad (2)$$

## IV. THE PROPOSED LFA-RTD MODEL

The proposed LFA-RTD model improves the basic LFA model for ITS data recovery [39, 40]. The spatio-temporal correlation is used as a regularization constraint to improve the recovery accuracy. Next, we introduce the proposed LFA-RTD model in detail.

### A. Spatio-Temporal Correlation

The spatio-temporal correlation of ITS has two basic characteristics, namely, spatial correlation and temporal correlation. The spatial correlation indicates that the data collected by vehicles near geographical locations are similar at every moment. Temporal correlation means that data collected from the same vehicle changes smoothly over time.

#### 1) Spatial correlation

In order to obtain spatial correlation, we constructed an undirected weighted vehicle connected graph $G = (V, E, W)$, $V$ ($|V|=M$) of them said vehicle vertices, $E$ is set, in which each edge said the link between the two vehicle, $W$ is weighted adjacency matrix, Where each entry $w_{i,j}$ in $i$-th row and $j$-th column represents the quantitative relationship between the $i$-th and $j$-th vertices.

Given $V$, we compute the distance matrix $P$ between vertices based on their coordinates, where each element $p_{i,i'}$ in $i$-th row and $i'$-th column represents the Euclidean distance between the $i$-th and $i'$-th vertices. According to $P$, we choose the closest $k$ vertices for each vertex and get $W$ as follows:

$$w_{i,i'} = \begin{cases} \dfrac{1}{p_{i,i'}}, & \text{if } i' \in Top_k(i) \\ 0, & \text{otherwise} \end{cases} \qquad (3)$$

Where $Top_k(i)$ denotes the closest $k$ vertices set of the $i$-th vertex. Let $L$ denote the Laplacian matrix of graph $G$, and $L$ is defined as follows:

$$L = \begin{bmatrix} sum(w_{1,\cdot}) & -w_{1,2} & \cdots & -w_{1,M-1} & -w_{1,M} \\ -w_{2,1} & sum(w_{2,\cdot}) & \cdots & -w_{2,M-1} & -w_{2,M} \\ \cdots & \cdots & \cdots & \cdots & \cdots \\ -w_{M,1} & -w_{M,2} & \cdots & -w_{M,M-1} & sum(w_{M,\cdot}) \end{bmatrix}_{M \times M} \qquad (4)$$

where $sum(w_{i,\cdot})$, $i \in \{1, 2, ..., M\}$, denotes the sum of $i$-th row of $W$ matrix. Then, the spatial correlation is captured by $L$.

#### 2) Differential correlation

To obtain temporal correlation, we make the data collected by the same vehicle in adjacent time periods similar to each other. To this end, we construct the time differential matrix $D$ as follows:

$$D = \begin{bmatrix} -1 & 0 & 0 & \cdots & 0 \\ 1 & -1 & 0 & \cdots & 0 \\ 0 & 1 & -1 & \cdots & 0 \\ \cdots & \cdots & \cdots & \cdots & \cdots \\ 0 & 0 & \cdots & 1 & -1 \\ 0 & 0 & \cdots & 0 & 1 \end{bmatrix}_{N \times N-1} \qquad (5)$$

#### 3) Regularization constraint of spatio-temporal correlation

After obtaining the Laplacian matrix $L$ and the time differential matrix $D$, we include them in (5) as regularization constraints:

$$\varepsilon(X,Y) = \frac{1}{2}\left\|J \circ (H - XY^{\mathrm{T}})\right\|_F^2 + \lambda(\|X\|_F^2 + \|Y\|_F^2) + \underbrace{\beta_1 \|LXY^{\mathrm{T}}\|_F^2}_{\text{Spatial smoothness constraint}} + \underbrace{\beta_2 \|XY^{\mathrm{T}}D\|_F^2}_{\text{Temporal smoothness constraint}} \quad (6)$$

where $\beta_1$ and $\beta_2$ are two hyper-parameters to affect the performance of spatial and temporal correlation, respectively.

### B. Model Optimization

We extend (6) to a single-element oriented form as follows [41, 42]:

$$\varepsilon(X,Y) = \underbrace{\sum_{y_{i,j} \in H_K} (h_{i,j} - x_{i,.} y_{j,.})^2}_{L_2\text{-norm-oriented Loss}} + \underbrace{\beta_1 \sum_{y_{i,j} \in H_K} \left((LXY^{\mathrm{T}})_{(i,j)}\right)^2}_{\text{Spatial smoothness constraint}}$$
$$+ \underbrace{\beta_2 \sum_{y_{i,j} \in H_K} \left((XY^{\mathrm{T}}D)_{(i,j)}\right)^2}_{\text{Temporal smoothness constraint}} + \underbrace{\lambda \sum_{y_{i,j} \in H_K} \left((x_{i,.})^2 + (y_{j,.})^2\right)}_{\text{Tikhonov regularization}}, \quad (7)$$

where $x_{i,.}$ and $y_{j,.}$ denote the $i$-th and the $j$-th row-vector of $X$ and $Y$, respectively. Then, considering the instant loss $\varepsilon_{i,j}$ of $\varepsilon(X, Y)$ on a single entry $h_{i,j}$, we have:

$$\varepsilon_{i,j} = (h_{i,j} - x_{i,.} y_{j,.})^2 + \beta_1 \left((LXY^{\mathrm{T}})_{(i,j)}\right)^2$$
$$+ \beta_2 \left((XY^{\mathrm{T}}D)_{(i,j)}\right)^2 + \lambda \left((x_{i,.})^2 + (y_{j,.})^2\right). \quad (8)$$

The optimization of (7) w.r.t. $x_{i,.}$ and $y_{j,.}$ can be achieved by the stochastic gradient descent (SGD) algorithm. Then, at the $t$-th iteration, we employ SGD to minimize (7) as follows:

$$\begin{cases} x_{i,.}^t = x_{i,.}^{t-1} - \eta \dfrac{\partial \varepsilon_{i,j}^{t-1}}{\partial x_{i,.}^{t-1}} \\ y_{j,.}^t = y_{j,.}^{t-1} - \eta \dfrac{\partial \varepsilon_{i,j}^{t-1}}{\partial y_{j,.}^{t-1}} \end{cases}, \quad (9)$$

where $x_{t-1\,i,.}$, $y_{t-1\,j,.}$, and $\varepsilon_{t-1\,i,j}$ denote the states of $x_{i,.}$, $y_{j,.}$, and $\varepsilon_{i,j}$ at the $(t-1)$-th iteration, and $\eta$ denotes the learning rate of SGD. Let $\Delta_{t\,i,j}^t = h_{i,j} - x_{t-1\,i,.} y_{t-1\,j,.}$ be the estimation error on a single entry $h_{i,j}$ at the $(t-1)$-th iteration. By combining (7) into (9), we have the updating rules of $x_{i,.}$ and $y_{j,.}$ on a single entry $h_{i,j}$ at the $t$-th iteration as follows:

$$\begin{cases} x_{i,.}^t = x_{i,.}^{t-1} - 2\eta\lambda x_{i,.}^{t-1} + 2\eta y_{j,.}^{t-1} \Delta_{i,j}^{t-1} \\ \quad - 2\beta_1 \eta \left((LL^{\mathrm{T}})_{i,.} X^{t-1} (y_{j,.}^{t-1})^{\mathrm{T}}\right) y_{j,.}^{t-1} \\ \quad - 2\beta_2 \eta \left(x_{i,.}^{t-1} (Y^{t-1})^{\mathrm{T}} (DD^{\mathrm{T}})_{.,j}\right) y_{j,.}^{t-1} \\ y_{j,.}^t = y_{j,.}^{t-1} + 2\eta x_{i,.}^{t-1} \Delta_{i,j}^{t-1} - 2\eta\lambda y_{j,.}^{t-1} \\ \quad - 2\beta_1 \eta \left((LL^{\mathrm{T}})_{i,.} X^{t-1} (y_{j,.}^{t-1})^{\mathrm{T}}\right) x_{i,.}^{t-1} \\ \quad - 2\beta_2 \eta \left(x_{i,.}^{t-1} (Y^{t-1})^{\mathrm{T}} (DD^{\mathrm{T}})_{.,j}\right) x_{i,.}^{t-1} \end{cases} \quad (10)$$

### C. Algorithm Design and Time Complexity

In (11), the spatio-temporal correlation has much computational burden. We consider the spatio-temporal correlation by every few entries to improve computational efficiency. To this end, we first remove the spatio-temporal correlation from (11) as follows:

$$\begin{cases} x_{i,.}^t = x_{i,.}^{t-1} + 2\eta y_{j,.}^{t-1} \Delta_{i,j}^{t-1} - 2\eta\lambda x_{i,.}^{t-1} \\ y_{j,.}^t = y_{j,.}^{t-1} + 2\eta x_{i,.}^{t-1} \Delta_{i,j}^{t-1} - 2\eta\lambda y_{j,.}^{t-1} \end{cases}. \quad (11)$$

Then, we adopt an index $\mu$ to control the merging of spatio-temporal correlations. Therefore, the algorithm we designed, LFA-RTD, is shown in Table II. Its time complexity is mainly determined by the maximum number of iterations, the number of known entries of the target sparse matrix, the number of calculated spatio-temporal correlations and the dimension of the latent factors. Thus, its time cost is $\Theta(T \times |H_K| \times k \times M \times N/\mu)$.

TABLE II. ALGORITHM LFA-RTD.

| Input: $L, H_K$; Output: $X, Y$. | |
|---|---|
| Steps | Operation |
| 1 | initialize $d, \lambda, \eta, T, \mu$ |
| 2 | initialize $X$ randomly |
| 3 | initialize $Y$ randomly |
| 4 | while $t \leq T$ && not converge |
| 5 |   for $j$=1 to $N$ // input data by column of $H$ |
| 6 |     Index=0 |
| 7 |     for each $h_{i,j}$ in $H_K$ |
| 8 |       if Index≠$\mu$ |
| 9 |         update $x^t_{i,\cdot}$ according to (11) |
| 10 |       else update $x^t_{i,\cdot}$ according to (10) |
| 11 |       end if |
| 12 |     end for |
| 13 |     Index= Index+1 |
| 14 |   end for |
| 15 |   for $i$=1 to $M$ // input data by row of $H$ |
| 16 |     Index=0 |
| 17 |     for each $h_{i,j}$ in $H_K$ |
| 18 |       if Index≠$\mu$ |
| 19 |         update $y^t_{j,\cdot}$ according to (11) |
| 20 |       else update $y^t_{j,\cdot}$ according to (10) |
| 21 |       end if |
| 22 |     end for |
| 23 |     Index= Index+1 |
| 24 |   end for |
| 25 |   $t=t+1$ |
| 26 | end while |

## V. EXPERIMENTS AND RESULTS

In the next experiments, we aim at answering the following research question(RQs):
- RQ.1. How does the performance compare to other models in terms of recovery?
- RQ.2. How do hyper-parameters influence the performance of the proposed LFA-RTD model?

*A. General Settings*

***Datasets.*** Two benchmark datasets are selected to conduct the experiments. They are real datasets generated by ITS, including Shanghai and Shenzhen. Table III summarizes their properties.

TABLE III. THE PROPERTIES OF EXPERIMENTAL DATASETS

| No. | Name | $M$ | $N$ | Time |
|---|---|---|---|---|
| D1 | Shenzhen | 664 | 720 | 2007-2-20 All day |
| D2 | Shanghai | 1000 | 720 | 2014-10-22 All day |

***Evaluation Metrics.*** To evaluate the accuracy of missing data recovery, we adopt the root mean squared error (RMSE) as the evaluation metrics. They are calculated as follows:

$$RMSE = \sqrt{\left(\sum_{y_{i,j} \in \Gamma} (y_{i,j} - \hat{y}_{i,j})^2\right) \Big/ |\Gamma|}, \quad (12)$$

where $\hat{y}_{i,j}$ denote the estimation of $y_{i,j}$ and $\Gamma$ denotes the testing set.

***Baselines.*** We compare the proposed LFA-RTD model with four state-of-the-art models. Table IV gives brief descriptions of these competitors.

***Experimental Details.*** In order to simulate the situation of missing data in ITS, we randomly selected part of the data from the complete data set as the training set and the rest as the test set. In the training set, we use half of the data to train the model and the other half to validate the performance to optimize the hyper-parameters. After finding the optimal hyper-parameters, we retrain the model on all the data in the training set. The maximum number of iterations is set to 3000. In addition, if the error difference between two consecutive iterations is less than $10^{-6}$, the training of the model is terminated. We repeated each test five times and reported the average results. All experiments were performed on a PC equipped with a 3.4 GHZ I7 CPU and 64 GB RAM.

TABLE IV. SUMMARY OF COMPARED MODELS

| Model | Description |
|---|---|

| | | | |
|---|---|---|---|
| ST-LRMA [43] | It jointly applies the global and local correlations to build an LRMA model for recovering the spatio-temporal signal. | | |
| BR-TVGS [44] | It is a batch reconstruction-based LRMA model for recovering time-varying graph signals. | | |
| LRDS [45] | It introduces the differential smooth prior of time-varying graph signals into the LRMA model for recovering the spatio-temporal signal. | | |
| FPCA [46] | It proposes fixed point and Bregman iterative algorithms for solving the nuclear norm minimization problem. | | |
| LFA-RTD | It is a latent feature analysis(LFA) method combined with spatial-temporal to recovery traffic data(RTD). | | |

## B. Performance Comparison (RQ.1)

In this set of experiments, we increased the sampling rate from 0.1 to 0.9 to compare the performance of all the models involved. The comparison results of each data set are recorded in Table V.

To make better use of these results, we performed win/loss statistic analyses, Wilcoxon signed rank test and Friedman test. Win/lose is to calculate how many cases LFA-RTD has lower/higher RMSE than each comparison model when the sampling rate increases from 0.1 to 0.9. Wilcoxon signed rank test is a non-parametric pairwise comparison method to test whether LFA-RTD has significantly better recovery accuracy than each comparison model by the significance level of P-value. Friedman test is to compare the performance of multiple models in multiple situations simultaneously by F-rank value. The lower the F-rank value, the higher the recovery accuracy.

Table V shows that when the sample rate rises, the RMSE drops. We also have the following significant discoveries:
- Most of the time, LFA-RTD achieves the lowest RMSE.
- At the significance threshold of 0.05, all P-values are less than 0.05, demonstrating that LFA-RTD has a significantly higher recovery accuracy than all evaluated models.
- LFA-RTD outperforms all other models in terms of F-rank, further demonstrating that it has the highest recovery accuracy on the test dataset.

Therefore, the experimental results demonstrate that LFA-RTD significantly outperforms the comparison models in recovering the missing data from ITS.

TABLE V. THE COMPARISON RESULTS OF RECOVERY ACCURACY WITH DIFFERENT SAMPLING RATES.

| Sampling rate | Datasets | ST-LRMA | BR-TVGS | LRDS | FPCA | LFA-RTD |
|---|---|---|---|---|---|---|
| 0.1 | D1 | 0.96023 | 0.89909 | 1.06125 | 1.18083 | 0.82026 |
|  | D2 | 0.74236 | 1.04887 | 0.58988 | 0.88180 | 0.10982 |
| 0.2 | D1 | 0.89486 | 1.27572 | 0.92710 | 1.33508 | 0.77478 |
|  | D2 | 0.65275 | 1.03980 | 0.81132 | 0.83659 | 0.10004 |
| 0.3 | D1 | 0.85993 | 0.82864 | 0.87251 | 1.21606 | 0.75236 |
|  | D2 | 0.60781 | 1.03518 | 0.68510 | 0.80993 | 0.10026 |
| 0.4 | D1 | 0.83839 | 0.76987 | 0.84443 | 1.48745 | 0.73721 |
|  | D2 | 0.58218 | 1.03576 | 0.68180 | 0.81019 | 0.09428 |
| 0.5 | D1 | 0.81890 | 0.74641 | 0.82316 | 1.28117 | 0.72785 |
|  | D2 | 0.56629 | 1.03513 | 0.60619 | 0.74140 | 0.08914 |
| 0.6 | D1 | 0.80527 | 0.73109 | 0.81011 | 0.93702 | 0.72337 |
|  | D2 | 0.55278 | 1.03548 | 0.58988 | 0.65392 | 0.09069 |
| 0.7 | D1 | 0.79138 | 0.71557● | 0.79647 | 0.89289 | 0.71671 |
|  | D2 | 0.54657 | 1.03465 | 0.58216 | 0.62220 | 0.08934 |
| 0.8 | D1 | 0.77780 | 0.70461● | 0.78371 | 0.86872 | 0.71929 |
|  | D2 | 0.54056 | 1.04092 | 0.57610 | 0.60550 | 0.08612 |
| 0.9 | D1 | 0.77070 | 0.69205● | 0.77709 | 0.85349 | 0.70757 |
|  | D2 | 0.52837 | 1.03155 | 0.56210 | 0.58792 | 0.08311 |
| Statistical Analysis | win/loss♦ | 18/0 | 15/3 | 18/0 | 18/0 | —— |
|  | F-rank* | 2.50000 | 3.44444 | 3.38889 | 4.50000 | **1.16667** |
|  | p-value | 0.00011 | 0.00040 | 0.00011 | 0.00011 | —— |

♦ The win/loss count of LFA-RTD has lower/higher RMSE than each comparison model.
★ The *p*-value of the Wilcoxon signed-ranks test, where the accepted hypothese with a significance level of 0.05 are highlighted.
\* The F-rank of the Friedman test, where a smaller F-rank value denotes a higher recovery accuracy.
● Indicates the RMSE of our LFA-RTD is higher than that of comparison model.

### C. Influence of hyper-parameters (RQ.2)

According to prior studies [38-41], a larger latent factor dimension makes an LFA model have a better representation learning ability. Fig. 2 records the results with different latent factor dimensions from 10 to 90. We see that RMSE keeps decreasing as $k$ increases, which also substantiates that a larger $k$ can improve LFA-RTD's recovery accuracy. However, the accuracy gain is not significant when $k$ increases over a certain threshold, i.e., 40 on D1, and the change of k in the D2 dataset has little change on RMSE. Besides, section IV shows that a larger $k$ increases the time complexity. Therefore, to balance the recovery accuracy and computational efficiency, we recommend setting $k$=40 in practical applications.

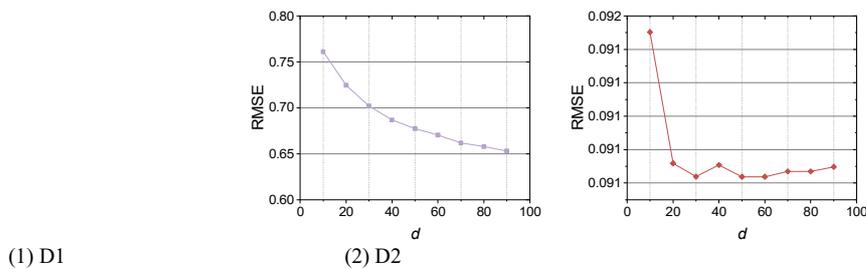

Fig.2. The RMSE of LFA-RTD as *k* increases from 10 to 90 on all the datasets.

| (1) D1 | (2) D2 |

## VI. CONCLUSIONS

Aiming at the problem of data recovery in Intelligent transportation systems (ITS), a spatio-temporal Signal recovery model based on Latent Feature Analysis (LFA) was proposed. The main idea is to incorporate spatio-temporal correlation into the LFA model as a regularization constraint to improve its recovery accuracy. Through this design, our LFA-RTD model has the advantages of high accuracy and robustness in traffic data recovery. In the future, we plan to make $β_1$ and $β_2$ adaptive through an intelligent optimization algorithm.

## VII. REFERENCES

[1] Q. Shang, Z. Yang, S. Gao, and D. Tan, "An Imputation Method for Missing Traffic Data Based on FCM Optimized by PSO-SVR," *Journal of Advanced Transportation*, vol. 2018, p. e2935248, 2018.
[2] Z. Li, S. Li and X. Luo, "An Overview of Calibration Technology of Industrial Robots," *IEEE/CAA J. Autom. Sinica*, vol. 8, no. 1, pp. 23-36, Jan. 2021.
[3] S. Tak, S. Woo, H. Yeo. "Data-driven imputation method for traffic data in sectional units of road links," *IEEE Transactions on Intelligent Transportation Systems*, vol. 17, no. 6, pp. 1762-1771, 2016.
[4] L. Hu, X. Pan, Z. Tan and X. Luo, "A Fast Fuzzy Clustering Algorithm for Complex Networks via a Generalized Momentum Method," *IEEE Transactions on Fuzzy Systems*, doi: 10.1109/TFUZZ.2021.3117442.
[5] L. Li, J. Zhang, Y. Wang et al. "Missing value imputation for traffic-related time series data based on a multi-view learning method," *IEEE Transactions on Intelligent Transportation Systems*, vol.20, no. 8, pp. 2933-2943, 2018.
[6] X. Shi, Q. He, X. Luo, Y. Bai and M. Shang, "Large-Scale and Scalable Latent Factor Analysis via Distributed Alternative Stochastic Gradient Descent for Recommender Systems," *IEEE Transactions on Big Data*, vol. 8, no. 2, pp. 420-431, 1 April 2022.
[7] X. Luo, Y. Zhou, Z. Liu, L. Hu and M. Zhou, "Generalized Nesterov's Acceleration-incorporated Non-negative and Adaptive Latent Factor Analysis," *IEEE Transactions on Services Computing*, doi: 10.1109/TSC.2021.3069108.
[8] D. Wu, Y. He, X. Luo and M. Zhou, "A Latent Factor Analysis-Based Approach to Online Sparse Streaming Feature Selection," *IEEE Transactions on Systems, Man, and Cybernetics: System*, doi: 10.1109/TSMC.2021.3096065.
[9] C. Xiang et al., "Edge computing-empowered large-scale traffic data recovery leveraging low-rank theory," *IEEE Transactions Network Science and Engineering*, vol. 7, no. 4, pp. 2205–2218, 2020.
[10] K. Qiu, X. Mao, X. Shen et al. "Time-varying graph signal reconstruction," *IEEE Journal of Selected Topics in Signal Processing*, vol.11, no. 6, pp. 870-883, 2017.
[11] L. Hu, S. Yang, X. Luo and M. Zhou, "An Algorithm of Inductively Identifying Clusters From Attributed Graphs," *IEEE Transactions on Big Data*, vol. 8, no. 2, pp. 523-534, 1 April 2022.
[12] X. Luo, Z. Liu, S. Li et al. "A fast non-negative latent factor model based on generalized momentum method," *IEEE Transactions on Systems, Man, and Cybernetics: Systems*, vol. 51, no.1, pp. 610-620, 2018.
[13] X. Luo, H. Wu, Z. Wang et al. "Non-Negative Latent Factor Model Based on $β$-Divergence for Recommender Systems," *IEEE Transactions on Systems, Man, and Cybernetics: Systems*, vol. 51, no. 8, pp. 4612-4623, Aug. 2021.
[14] X. Luo, Y. Zhou, Z. Liu, and M. Zhou, "Fast and Accurate Non-negative Latent Factor Analysis on High-dimensional and Sparse Matrices in Recommender Systems," *IEEE Transactions on Knowledge and Data Engineering*, pp. 1–1, 2021.
[15] X Luo, MC Zhou, S Li, Y Xia, ZH You, QS Zhu, H Leung, "Incorporation of Efficient Second-Order Solvers Into Latent Factor Models for Accurate Prediction of Missing QoS Data," *IEEE Transactions on Cybernetics*, vol. 48, no. 4, pp. 1216–1228, Apr. 2018.
[16] X. Luo, H. Wu, H. Yuan and M. Zhou, "Temporal Pattern-Aware QoS Prediction via Biased Non-Negative Latent Factorization of Tensors," *IEEE Transactions on Cybernetics*, vol. 50, no. 5, pp. 1798-1809, May 2020.
[17] D. Wu, P. Zhang, Y. He and X. Luo, "A Double-Space and Double-Norm Ensembled Latent Factor Model for Highly Accurate Web Service QoS Prediction," *IEEE Transactions on Services Computing*, doi: 10.1109/TSC.2022.3178543.
[18] S. Zhai, G. Li, G. Zhang, and Z. Qi, "Spatio-temporal signal recovery under diffusion-induced smoothness and temporal correlation priors," *IET Signal Processing*, vol. 16, no. 2, pp. 157–169, 2022.
[19] P. Wu, L. Xu, and Z. Huang, "Imputation Methods Used in Missing Traffic Data: A Literature Review," *Artificial Intelligence Algorithms and Applications*, pp. 662–677, 2020.
[20] Y. Koren, R. Bell and C. Volinsky, "Matrix Factorization Techniques for Recommender Systems," *Computer*, vol. 42, no. 8, pp. 30-37, Aug. 2009.
[21] X. Luo, M. Zhou, S. Li, Z. You, Y. Xia and Q. Zhu, "A Nonnegative Latent Factor Model for Large-Scale Sparse Matrices in Recommender Systems via Alternating Direction Method," *IEEE Transactions on Neural Networks and Learning Systems*, vol. 27, no. 3, pp. 579-592, March 2016.
[22] H. Wu, Z. Zhang, K. Yue, B. Zhang, J. He, and L. Sun, "Dual-regularized matrix factorization with deep neural networks for recommender systems," *Knowledge-Based Systems*, vol. 145, pp. 46–58, Apr. 2018.
[23] D. Wu, Q. He, X. Luo, M. Shang, Y. He and G. Wang, "A Posterior-Neighborhood-Regularized Latent Factor Model for Highly Accurate Web Service QoS Prediction," *IEEE Transactions on Services Computing*, vol. 15, no. 2, pp. 793-805, 1 March-April 2022.
[24] X. Luo, H. Wu and Z. Li, "NeuLFT: A Novel Approach to Nonlinear Canonical Polyadic Decomposition on High-Dimensional Incomplete Tensors," *IEEE Transactions on Knowledge and Data Engineering*, doi: 10.1109/TKDE.2022.3176466.
[25] X. Luo, Y. Yuan, S. Chen, N. Zeng and Z. Wang, "Position-Transitional Particle Swarm Optimization-Incorporated Latent Factor Analysis," *IEEE Transactions on Knowledge and Data Engineering*, vol. 34, no. 8, pp. 3958-3970, 2022.
[26] D. Wu, Q. He, X. Luo, M. Shang, Y. He and G. Wang, "A Posterior-Neighborhood-Regularized Latent Factor Model for Highly Accurate Web Service QoS Prediction," in IEEE Transactions on Services Computing, vol. 15, no. 2, pp. 793-805, 2022.
[27] D. Wu, Q. He, X. Luo, M. Shang, Y. He and G. Wang, "A Posterior-Neighborhood-Regularized Latent Factor Model for Highly Accurate Web Service QoS Prediction," *IEEE Transactions on Services Computing*, vol. 15, no. 2, pp. 793-805, 2022.
[28] B. L. Smith, W. T. Scherer, and J. H. Conklin, "Exploring Imputation Techniques for Missing Data in Transportation Management Systems," *Transportation Research Record*, vol. 1836, no. 1, pp. 132–142, Jan. 2003.

[29] D. Ni and J. D. Leonard, "Markov Chain Monte Carlo Multiple Imputation Using Bayesian Networks for Incomplete Intelligent Transportation Systems Data," Transportation Research Record, vol. 1935, no. 1, pp. 57–67, Jan. 2005

[30] L. Qu, L. Li, Y. Zhang, and J. Hu, "PPCA-Based Missing Data Imputation for Traffic Flow Volume: A Systematical Approach," *IEEE Transactions on Intelligent Transportation Systems*, vol. 10, no. 3, pp. 512–522, Sep. 2009.

[31] I. B. Aydilek and A. Arslan, "A hybrid method for imputation of missing values using optimized fuzzy c-means with support vector regression and a genetic algorithm," *Information Sciences*, vol. 233, pp. 25–35, Jun. 2013.

[32] S. Tak, S. Woo, and H. Yeo, "Data-Driven Imputation Method for Traffic Data in Sectional Units of Road Links," *IEEE Transactions on Intelligent Transportation Systems*, vol. 17, no. 6, pp. 1762–1771, Jun. 2016.

[33] I. Laña, I. (Iñaki) Olabarrieta, M. Vélez, and J. Del Ser, "On the imputation of missing data for road traffic forecasting: New insights and novel techniques," *Transportation Research Part C: Emerging Technologies*, vol. 90, pp. 18–33, May 2018.

[34] M. Berglund, T. Raiko, M. Honkala, L. Kärkkäinen, A. Vetek, and J. T. Karhunen, "Bidirectional Recurrent Neural Networks as Generative Models,"*Advances in Neural Information Processing Systems*, 2015, vol. 28.

[35] Y. Duan, Y. Lv, Y.-L. Liu, and F.-Y. Wang, "An efficient realization of deep learning for traffic data imputation," *Transportation Research Part C: Emerging Technologies*, vol. 72, pp. 168–181, Nov. 2016.

[36] L. Li, J. Zhang, Y. Wang, and B. Ran, "Missing Value Imputation for Traffic-Related Time Series Data Based on a Multi-View Learning Method," *IEEE Transactions on Intelligent Transportation Systems*, vol. 20, no. 8, pp. 2933–2943, 2019.

[37] R. Asadi and A. Regan, "A convolution recurrent autoencoder for spatio-temporal missing data imputation." *arXiv*, Apr. 28, 2019. doi: 10.48550/arXiv.1904.12413.

[38] D. Wu, M. Shang, X. Luo, and Z. Wang, "An $L_1$-and-$L_2$-Norm-Oriented Latent Factor Model for Recommender Systems," *IEEE Transactions on Neural Networks and Learning Systems*, pp. 1–14, 2021, doi: 10.1109/TNNLS.2021.3071392.

[39] D. Wu and X. Luo, "Robust Latent Factor Analysis for Precise Representation of High-Dimensional and Sparse Data," *IEEE/CAA Journal of Automatica Sinica*, vol. 8, no. 4, pp. 796–805, Apr. 2021.

[40] D. Wu, X. Luo, M. Shang, Y. He, G. Wang, and X. Wu, "A Data-Characteristic-Aware Latent Factor Model for Web Services QoS Prediction," *IEEE Transactions on Knowledge and Data Engineering*, vol. 34, no. 6, pp. 2525–2538, Jun. 2022.

[41] D. Wu, X. Luo, M. Shang, Y. He, G. Wang, and M. Zhou, "A Deep Latent Factor Model for High-Dimensional and Sparse Matrices in Recommender Systems," *IEEE Transactions on Systems, Man, and Cybernetics: Systems*, vol. 51, no. 7, pp. 4285–4296, Jul. 2021.

[42] L. Hu, X. Yuan, X. Liu, S. Xiong, and X. Luo, "Efficiently Detecting Protein Complexes from Protein Interaction Networks via Alternating Direction Method of Multipliers," IEEE/ACM Transactions on Computational Biology and Bioinformatics, vol. 16, no. 6, pp. 1922–1935, 2019.

[43] X. Piao, Y. Hu, Y. Sun, B. Yin, and J. Gao, "Correlated Spatio-Temporal Data Collection in Wireless Sensor Networks Based on Low-Rank Matrix Approximation and Optimized Node Sampling," *Sensors*, vol. 14, no. 12, pp. 23137–23158, 2014.

[44] K. Qiu, X. Mao, X. Shen, X. Wang, T. Li, and Y. Gu, "Time-varying graph signal reconstruction," I*EEE J. Sel. Topics Signal Process*., vol. 11, no. 6,pp. 870–883, Sep. 2017.

[45] X. Mao, K. Qiu, T. Li, and Y. Gu, "Spatio-Temporal Signal Recovery Based on Low Rank and Differential Smoothness," *IEEE Trans. Signal Process.*, vol. 66, no. 23, pp. 6281–6296, 2018.

[46] S. Ma, D. Goldfarb, and L. Chen, "Fixed point and Bregman iterative methods for matrix rank minimization," Math. Program., vol. 128, no. 1, pp. 321–353, Jun. 2011.